Tuning Superconductivity and Charge Density Wave Order in 2H-TaSe$_2$ by Modulating the Van Hove Singularity

Mukhtar Lawan Adam[1,2*], Ibrahim Buba Garba[3,4], Abba Alhaji Bala[5], Abdulsalam Aji Suleiman[6*], Sulaiman Muhammad Gana[2], and Faisal Lawan Adam[2]

[1]Materials Interfaces Center, Shenzhen Institute of Advanced Technology, Chinese Academy of Sciences, Shenzhen 518055, Guangdong, PR China.
[2]Physics Department, Bayero University, Kano, 700231, Nigeria.
[3]Sorbonne Université, CNRS, IMPMC, 4 Place Jussieu, 75252 Paris, France.
[4]Department of Physics, Federal University Gashua, 671106, Gashua, Nigeria.
[5]Physics Department, Federal University Dutse, Jigawa, Nigeria.
[6] Institute of Materials Science and Nanotechnology, Bilkent University UNAM, Ankara 06800, Turkey.

[*]Corresponding authors: mladam@siat.ac.cn; Ibrahim.garba@sorbonne-universite.fr; abdulsalam@unam.bilkent.edu.tr

**Abstract**

Tantalum diselenide (TaSe$_2$) is an exciting material that hosts charge density wave order (CDW) and superconductivity. Thus, providing a playing field for examining the interactions of fundamental electronic quantum states in materials. Recent research has proposed that the intrinsic quantum electronic state in the TaSe$_2$ lattice could be improved by aligning the Van Hove singularity (VHs) with the Fermi level. In this study, we attempt to tune the VHs in TaSe$_2$ to align them within the vicinity of the Fermi level via electron doping by chemically substituting Pt for Ta atoms. On investigating the band structure of Pt$_{0.2}$Ta$_{0.8}$Se$_2$, the electron doping brought the VHs closer to the Fermi level vicinity around the **K** high symmetry point. As a result, the CDW state in pristine TaSe$_2$ is suppressed in the TaSe$_2$ doped system while also hosting an enhanced superconducting temperature (T$_c$) of ~2.7 K. These observations provide insight into ways to leverage the VHs in materials to tune their electronic properties.

**Introduction**

The presence of Van Hove singularities (VHs) around the vicinity of the Fermi surface (FS) of materials is observed to be accompanied by the emergence of low-temperature phenomena in materials [1–3]. These electronic instabilities at the FS vicinity introduce delocalized electrons into the 'Fermi sea', creating low-energy excitations, thus significantly impacting the overall electronic properties of materials. This has led to the formation of exotic quantum states such as superconductivity, ferromagnetism, and spin-density wave states in kagome metals, graphene, sulfur hydrides, and transition metal dichalcogenides [4–13]. Due to the emergence of these exotic phenomena, several routes, such as doping and pressure, have been exploited to tune the VHs in materials into the FS vicinity [9,14–17]. The VHs is a universal electronic artifact that forms a saddle-like feature in the electronic band structure of quasi-two dimensional materials [18,19]. At the VHs, the density of states (DOS) diverges logarithmically to form highly localized electronic states [15,20]. On tuning the VHs to the Fermi vicinity, the FS of materials undergoes a Lifshitz transition modifying their overall electronic structure while preserving the structural symmetry [21–23].

Chemical doping is one more experimentally accessible route for inducing quantum material changes. Quantum materials with VHs existing above/below the Fermi level can be tuned by electron/hole doping to fall within the as-desired position to tune existing or induce new electronic quantum phenomena [4,17,24]. Bulk $TaSe_2$ is an intriguing quantum material co-hosting superconductivity at $T_c$ ~ 0.1 K and Charge density wave (CDW) order transition at ~110K [25]. Furthermore, it hosts VHs around ~ 0.5eV above the Fermi level along the ***K-G*** high symmetry line [26]. Interestingly, the CDW order is known to compete with superconductivity, decreasing the DOS at the Fermi level. Thus, it can suppress the formation of Cooper pairs and the transition temperature.

Moreover, the existence of Fermi surface nesting in 2H-TaSe$_2$ has also been attributed to the formation of a CDW order [25]. In essence, the complete suppression of one of these competing states, i.e., the CDW, can enhance the one, i.e., the superconductivity. Taking advantage of the VHs in 2H-TaSe$_2$ and careful electron doping to tune it into the FS vicinity, the CDW order can be suppressed [27,28]. These will enhance the DOS around the Fermi to facilitate the formation of Cooper pairs therein.

In this work, we attempt to introduce electrons into 2H-TaSe$_2$ through the controlled doping of Pt atoms, and these could lower the chemical potential pushing the VHs into the Fermi vicinity. The doping of Pt atoms into TaSe$_2$ substitutes the Ta atoms, as confirmed through X-ray diffraction and DFT calculations. At 0.2% doped Pt, the VHs feature aligned within the vicinity of the Fermi level. An enhanced superconducting transition temperature of ~2.7 K and complete suppression of the CDW phase as observed in the pristine TaSe$_2$.

**Experimental Details**

Single crystals of Pt$_x$Ta$_{1-x}$Se$_2$ (x = 0, 0.05, 0.1, 0.2) were synthesized using the chemical vapor transport method. Pt, Ta, and Se powders were combined with a varying amount of Iodine to serve as a transport agent, and the mixture was then sealed in an air-evacuated ampoule tube. The sealed tubes were then placed at room temperature in a two-zone furnace and heated for five days while maintaining the furnace at 900 ºC in the evaporation zone and 800 ºC in the growth zone. The furnace was cooled to room temperature after five days. High-quality single crystals were obtained at the end of the ampoule. The crystal's diffraction was measured using a Panalytical X-ray diffractometer with Kα radiation. Transport properties measurements at low were carried out using a Quantum Design physical property measurement system at the Hefei center for instruments, university of science and technology, China. All First principle calculations were carried out using the Quantum espresso density functional theory package to gain insight into the electronic properties of the compounds.

**Results and discussions**

According to our experimental and theoretical findings, Pt atoms doped into $TaSe_2$ introduce electron-like charge carriers into the compound, as seen by our XPS data, **Figure 2**. The presence of such charge carriers in semimetal/metallic materials while hosting quantum electronic states behavior such as charge density wave/superconductivity is known to modulate such states. In our case, carrier doping in $TaSe_2$ resulted in the suppression of the charge density wave states while enhancing the superconducting transition temperature, as demonstrated by transport properties measurement carried out at low temperatures.

Single crystals of the compounds $Pt_xTa_{1-x}Se_2$ (x = 0, 0.05, 0.1, and 0.2) were synthesized using the chemical vapor transport method. As illustrated in Figure 1(a), X-ray diffraction of the as-synthesized crystals and DFT calculations revealed that the Pt atoms substitution into the Ta atoms position was more energetically favorable than intercalation between the layers of the $TaSe_2$. ). The x-ray diffraction of the single crystals, as shown in **Figure 1(b)**, matches that of the pristine $TaSe_2$ with the space group of $P6_3/mmc$ without a shift in the peaks to lower degrees, indicating lattice expansion.

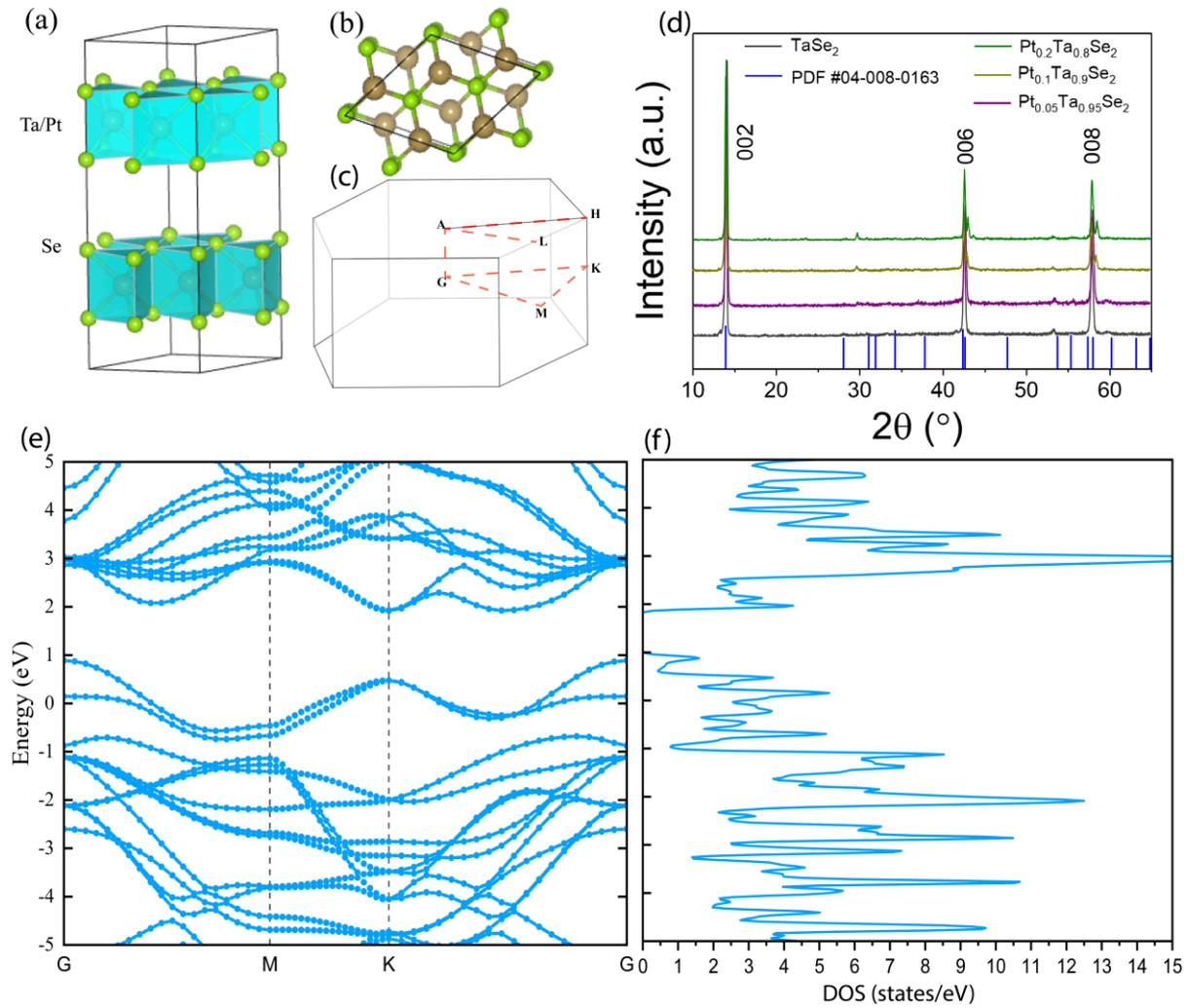

**Figure 1** **(a)** Side view **(b)** Top view of the crystal structure of Pt-doped $TaSe_2$. **(c)** The Brillion zone indicating the high symmetry points **(d)** Directly compare the X-ray diffraction (XRD) of pristine $TaSe_2$ and Pt-doped $TaSe_2$. **(e)** the band structure and **(f)** density of the state of 2H-$TaSe_2$.

Further on, we characterized the sample using X-ray photoelectron spectroscopy to investigate the role of Pt atoms in the compound and ascertain their chemical state. The XPS spectra of pristine $TaSe_2$ and $Pt_{0.2}Ta_{0.8}Se_2$ are shown in **Figure 2**. The XPS spectra of the Ta 4f region of TaSe$_2$ and Pt$_{0.2}$Ta$_{0.8}$Se$_2$, Show a slight shift from the latter to lower binding energy, indicating electron donation from Pt to the TaSe$_2$ and a formation of bonds with the Se atoms.

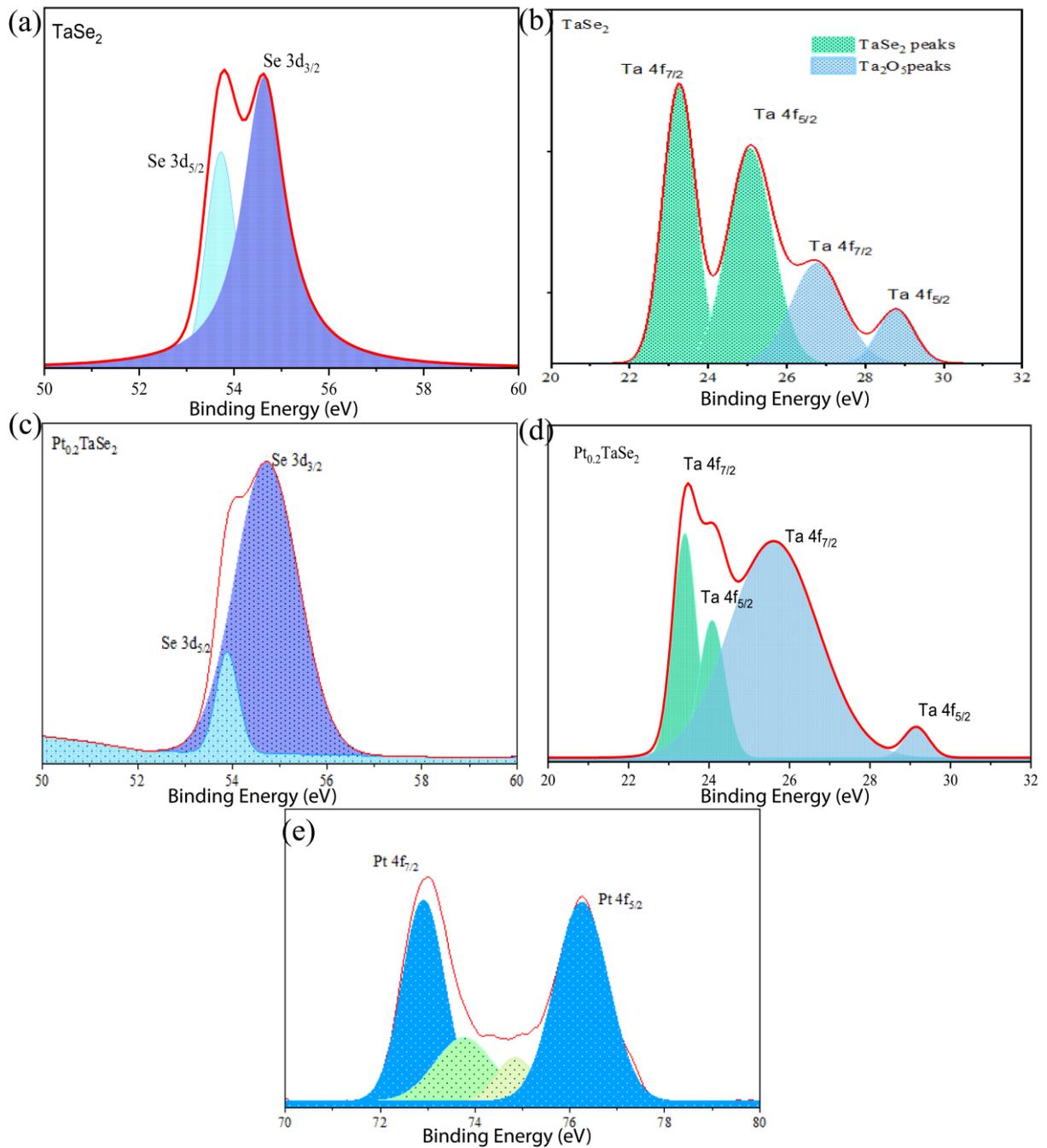

**Figure 2** The XPS spectra of **(a)** Se 3d region of TaSe$_2$ **(b)** Ta 4f region of TaSe$_2$ **(c)** Se 3d region of Pt$_{0.2}$Ta$_{0.8}$Se$_2$ **(d)** Ta 4f region of Pt$_{0.2}$Ta$_{0.8}$Se$_2$ **(e)** Pt 4f region of Pt$_{0.2}$Ta$_{0.8}$Se$_2$.

Previous research has shown that electron/hole doping into the TaSe$_2$ modifies intrinsic electronic properties [28–30]. The calculated bulk band structure of the Pt$_x$Ta$_{1-x}$Se$_2$ (x = 0.1 and

0.2), as shown in **Figure 3**, was calculated using the Wannier tool to distinguish between the compounds bulk and topological surface states easily, as well as the VHs features modulations. Neither of the compounds was observed to host non-trivial topological surface states from the bulk band structures in **Figures 3(a)** and **(b)**. However, there is significant modulation of the band structures of the compounds with the increase of Pt concentration from 0.1% to 0.2%. In comparison, as observed, the conduction bands show a downward shift in can be observed.

One of the most critical observations in the Fermi surface map of CDW materials like $TaSe_2$ is the presence of Fermi surface nesting [28]. At the normal phase of $TaSe_2$, its fermi surface is known to host hole-like pockets centered around **G** and **K** points, with "dog-bones" shaped electron pockets around the **M** point, [31,32] as shown in **Figure 3(c)**. However, at a commensurate CDW state with a 3x3 state, the Fermi surface undergoes a reconstruction, and the "dog bone" around **M** moves to the **G** point, while the hole-like pockets around the **K** degenerate twice to form concentric pockets [27,29,31–34]. **Figure 3(c)** shows that from the Fermi surface plots of $Pt_{0.1}TaSe_2$, there is almost an identical observation with the commensurate CDW Fermi surface of $TaSe_2$, indicating a possible presence of a CDW state. However, at its **K** point, electron-like pockets appear due to Pt doping.

Moreover, in an ARPES analysis of the CDW state suppression in $Pd_xTaSe_2$ [28], a topological Lifshitz transition is observed from electron-like pockets in the fermi surfaces as observed in the $TaSe_2$ to two hole-like pockets [35]. Accompanied by a sort of dismantling of the "dog bone," Fermi surfaces caused by the slight separation between the two hole-like pockets. Interestingly, these two hole-like pockets manifested a van Hove singularity of the saddle-point type, as observed in the Fermi surface map.

As Figure 3 (d) shows, two degenerate hole-like pockets appear at points **K** and **G** as the concentration of Pt increases from 0.1% to 0.2%. Moreover, these two hole-like pockets corroborate with a saddle-point van Hove singularity feature, as seen in **Figure 3(b)** around the **K** high symmetry point. Also, the "dog-bones" like the Fermi surface, as observed in $Pt_{0.1}TaSe_2$ around the **M** point, are eliminated in $Pt_{0.2}Ta_{0.8}Se_2$. Based on these findings, a change phase from a CDW material to a metal in $Pt_{0.2}TaSe_2$ can be said to have occurred.

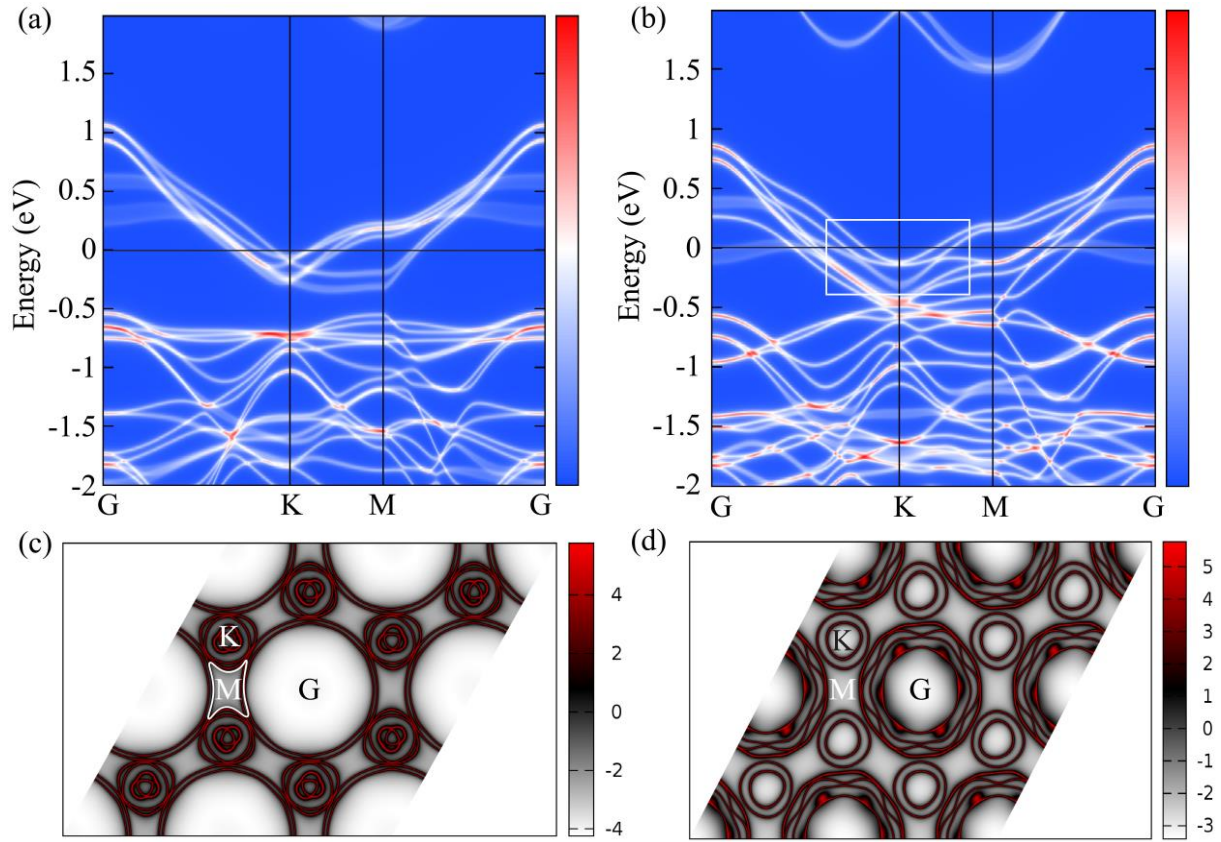

Figure 3 The band structure of (a) $Pt_{0.1}Ta_{0.9}Se_2$ (b) $Pt_{0.2}Ta_{0.8}Se_2$. The Fermi surface of (c) $Pt_{0.1}Ta_{0.9}Se_2$ (d) $Pt_{0.2}Ta_{0.8}Se_2$

The appearance of saddle-point Van Hove singularities close to the fermi level in materials amplifies the electron correlation, resulting in a CDW or superconductivity at low temperatures [4,36–38]. The transport measurements of the as-synthesized $Pt_xTa_{1-x}Se_2$ (x = 0.05, 0.1, 0.2) crystals at low temperatures were carried out to explore their exotic electronic properties. **Figures 4(a)** shows the temperature-dependent resistivity plot of the doped $TaSe_2$ crystals, where each of the $Pt_xTa_{1-x}Se_2$ (x = 0.05, 0.1) compounds display a typical metallic, with a distinct hump-like transition around ~100K, indicating the presence of a charge density wave behavior. The $Pt_xTa_{1-x}Se_2$ (x = 0.05, 0.1) compounds bearly superconduct. Interestingly, as demonstrated in **Figure 4(b)** $Pt_{0.2}Ta_{0.8}Se_2$, the CDW is completely suppressed, and hosts superconductivity at a transition temperature of ~2.7 K. At lower temperatures, **Figure 4(c)** displays the temperature-dependent resistive transition curves under the application of different magnetic fields, the response of superconducting transition temperature of $Pt_{0.2}Ta_{0.8}Se_2$ to an applied external magnetic field indicates a typical behavior of a superconductor. The superconducting temperature transition temperature broadens and shifts towards lower temperatures by increasing the fields due to a field-induced pair-breaking effect. Typical of superconducting materials

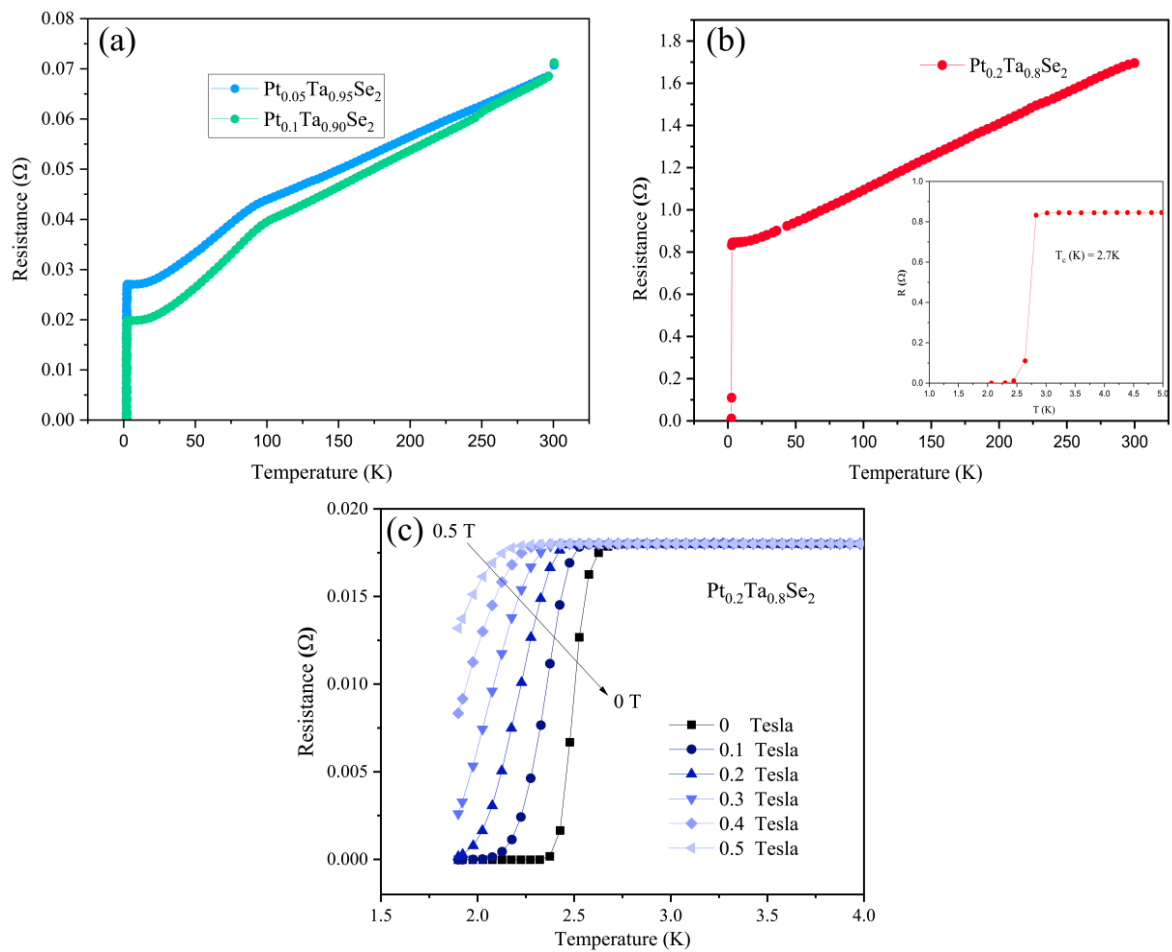

Figure 4 (a) the plot of resistivity against the temperature of $Pt_{0.05}Ta_{0.95}Se_2$ and $Pt_{0.1}TaSe_2$, indicating the charge density wave (CDW) transition. (b) the plot of resistivity against the temperature of $Pt_{0.2}Ta_{0.8}Se_2$ (inset showing the superconducting transition temperature). (c) Resistivity against temperature plot under a varying magnetic field of $Pt_{0.2}Ta_{0.8}Se_2$.

**Conclusion**

In conclusion, we have successfully synthesized and analyzed single crystals of $Pt_xTa_{1-x}Se_2$ (x = 0, 0.05, 0.1, and 0.2) and characterized them in terms of their physical, electronic structure, and chemical properties. The $Pt_{0.2}Ta_{0.8}Se_2$ superconductivity, in particular, is enhanced with a transition temperature of ~2.7 K while the CDW-ordering state is completely suppressed. According to the XPS investigations, the chemical doping of Pt atoms in $TaSe_2$ induces electron charge into the system. Our analyses revealed that electron doping in $TaSe_2$ can result in the emergence of saddle-point Van Hove singularities around the Fermi level, which can mediate exotic quantum electronic states therein.

**ACKNOWLEDGEMENTS**

M. L. A. acknowledges the China postdoctoral foundation, Bayero University, Kano,

National Synchrotron Radiation Laboratory, Hefei. I.B.G. acknowledges PTDF Fund.

**AUTHOR CONTRIBUTIONS**

M.L.A. and I.B.G. conceived the idea, supervised the project, and performed the experiments and calculations. M.L.A., A.A.S., and S.M.G. prepared the draft. All the authors participated in discussing the results and revising the manuscript.

**DECLARATION OF INTERESTS**

The authors declare no competing interests.